# ChemDuino-Calorimetry to Determine the Enthalpy Change of Neutralization of an Acid−Base Reaction: Making a Familiar Experiment "Greener"

N Krisnu Prabowo, M Paristiowati[a)], Irwanto, Afrizal, and Yusmaniar

*Department of Chemistry Education, Faculty of Mathematics and Natural Science, Universitas Negeri Jakarta, Jakarta, Indonesia*

[a)] Corresponding author: maria.paristiowati@unj.ac.id

**Abstract.** In thermochemistry experiments, a large volume and relatively high concentration of chemicals are often used to get a significant temperature change using an analogue thermometer. A 'greener' experimental protocol is developed as an alternative using modern microcontroller boards, ChemDuino. The aim of this research is to develop a low-cost and pocket-sized prototype for measuring temperature of solutions using Arduino Uno R3 DIP and DS18B20 sensor based on green chemistry and DIY (do-it –yourself) methods. The construction of ChemDuino for temperature sensing is made in the simplest manner for individuals with little knowledge of electronics. The details of all components are stipulated in this paper. This research was conducted in three phases, which includes preliminary, prototype development, and assessment phase. The device is used to determine the enthalpy change of neutralization (ΔH) of NaOH and HCl solution. The enthalpy change of neutralization, kJ mol$^{-1}$ at 25.0 °C from literature is −57.13 kJ mol$^{-1}$, whereas the enthalpy change of neutralization, kJ mol$^{-1}$ at 25.0 °C from the experiment using ChemDuino-Calorimetry is −56.87 (± 1.9) kJ mol$^{-1}$ (average of seven determinations and estimated standard deviation). The device has successfully measured the temperature change of the reaction at a relatively lower concentration. ChemDuino-Calorimetry has a great potential, because of its reliability and accuracy in measurements, inexpensive setup, and interconnectivity.

## INTRODUCTION

Arduino is one of the most well-known microcontrollers in the market today [1]. Various kinds of Arduino are available in the market around the world [2]. They are reasonably cost-effective, user-friendly, and readily available in a starter kit, including sensors, cables, breadboard, and other electronic components [3]. Some Arduino-integrated hardware have passed certain industrial certification for safety standards. The name "Arduino" came from a place in Italy, where the project's founders used to gather. Arduino opens many possibilities in its application since it can be coupled with multiple sensors, servo/motors, internet/in-built Wi-Fi, camera, smartphones, and computers via the Wiring interface [4]. It allows the Arduino to communicate with other electronic parts connected to it. An open-source tool called Arduino IDE (Integrated Development Environment), which is compatible with common operating systems such as Windows, Linux, and Mac, can be used to write code to the Arduino from a computer [5].

The integration of Arduino microcontroller for chemistry applications has evolved rapidly in the past five years. There are endless possibilities when using Arduino in chemistry [6]. The blend of the word chemistry and Arduino, ChemDuino, was first introduced in 2015 by Štěpánka Kubínová and Jan Šlégr [7]. ChemDuino is the standard implementation or praxis of the Arduino hardware and software in chemistry education. Many researchers have successfully developed and incorporated it to sense water hardness [8], to control sampling and record data (data acquisition devices) [9], to detect hazardous and toxic pollutants in a liquid or solution phase [10, 11, 12, 13] and gas phase [14, 15, 16], to detect pH of various solutions [17], to control a power switch in a solar cell system [18], and to monitor chemical experiments in real-time [19]. The ChemDuino-based learning environment has been reported to support online laboratory sessions and a hybrid virtual-physical environment [20], stimulate problem-solving and computational skills, which are connected to the significant increase of the student's attitudes towards environmental issues [21], be able to change learning difficulties into the agents of learning for visually impaired chemistry students [22], and increase motivation and engagement [23]. A fully-supported Arduino society with interdisciplinary subjects is reported to affect creativity in science positively [24]. However, advanced programming and a complex setup might hinder beginners from even looking for ChemDuino as an alternative solution for chemistry problems. Hence, in this paper, we are using the perspective of individuals with little knowledge of electronics.

In 2017, Bopegedera and Nishanthi implemented green chemistry principles in calorimetry by changing styrofoam cups with paper cups [25]. Styrofoam is a non-biodegradable material. Therefore, many countries worldwide have banned its use [26]. The study aimed to measure the enthalpy change of neutralization of an acid-base reaction using a (Do-It-Yourself) DIY calorimeter with paper cups. It is reported that three-layered paper cups provide comparable results with those from the styrofoam cup calorimeter. The maximum temperatures of the exothermic acid-base reactions are challenging to measure using an analogue thermometer. In 2020, Josu Lopez-Gazpio and Inigo Lopez-Gazpio published a paper utilising the ChemDuino-calorimeter to calculate the enthalpy change of neutralization of hydrochloric acid and sodium hydroxide solution. However, the volumes of the acid and base solutions used are relatively large, with a total volume of 600 cm$^3$. The concentrations of both acid and base used are 1 mol dm$^{-3}$ [27]. The green chemistry aspects might be overlooked. Clearly, a research gap needs to be filled between green chemistry aspects and ChemDuino implementation for the topic of thermochemistry. Therefore, we aim to develop a low-cost and pocket-sized prototype using ChemDuino as an alternative to measure the enthalpy change of neutralization of an acid-base reaction.

## MATERIALS AND METHODS

Sodium hydroxide, hydrochloric acid fuming 37%, pH indicator, and oxalic acid dehydrate were analytical-grade reagents from Merck (Germany). Distilled water was supplied by a local chemistry store near the school area. 0.100 mol dm$^{-3}$ sodium hydroxide and 0.100 mol dm$^{-3}$ hydrochloric acid solutions were used for the experiments. Standardization of sodium hydroxide solution was done with the oxalic acid solution.

ChemDuino was assembled from Arduino Uno R3 DIP, waterproof temperature sensor DS18B20, breadboard, resistor 10k, 4 male-to-female jumper cables, 3 male-to-male jumper cables, LCD 16x2 with soldered I2C module, and a USB connector. The sensor DS18B20 is manufactured by Semiconductor Corp, USA [28]. The sensor is 91 cm long and has a diameter of 4 mm. It is impact-resistant and robust waterproof. It is inexpensive and readily available with the wiring attached to the pin of the electronic sensor. The other brands of waterproof temperature sensors are usually sold without wiring (pin sensor only), making it challenging for beginners. The DS18B20 sensor was also used in health-monitoring system and biology [29, 30, 31], providing a reliable spectrum of applications.

The installation of codes was required in Arduino. It was done with Arduino IDE 1.8.16 [5]. All electronic components were purchased from a local robotic store. The DIY calorimeter was prepared from three paper cups and a tumbler from a local waste. They were washed, cleaned, and dried in a desiccator. The three paper cups fit perfectly into the tumbler, as depicted in Figure 2 (b). The prototype of ChemDuino was built first before the coding installation. The overall process covers three phases, which includes preliminary, prototype development, and assessment phase, shown in Figure 1. The stages were adapted from Plomp development model [32]. The circuit diagram describing the connection between the electronic parts can be seen in Figure 2 (a). It is an explanatory illustration exhibiting how the components are assembled. An open-source tool, Fritzing, was used to design this electronic prototype [33]. The cables from the temperature sensor, DS18B20, were soldered to fit firmly into the breadboard. All electronic parts were put inside a small box with the sensor and USB cable sticking out. Arduino IDE was used to write and install the codes into the microcontroller so that it could function. The codes and steps to use Arduino IDE were given in the additional information (supplementary content) section. Two codes were developed after one another. The code version 1.0 was much simpler code. This code was tested and evaluated. The second code, version 1.1, was written after that.

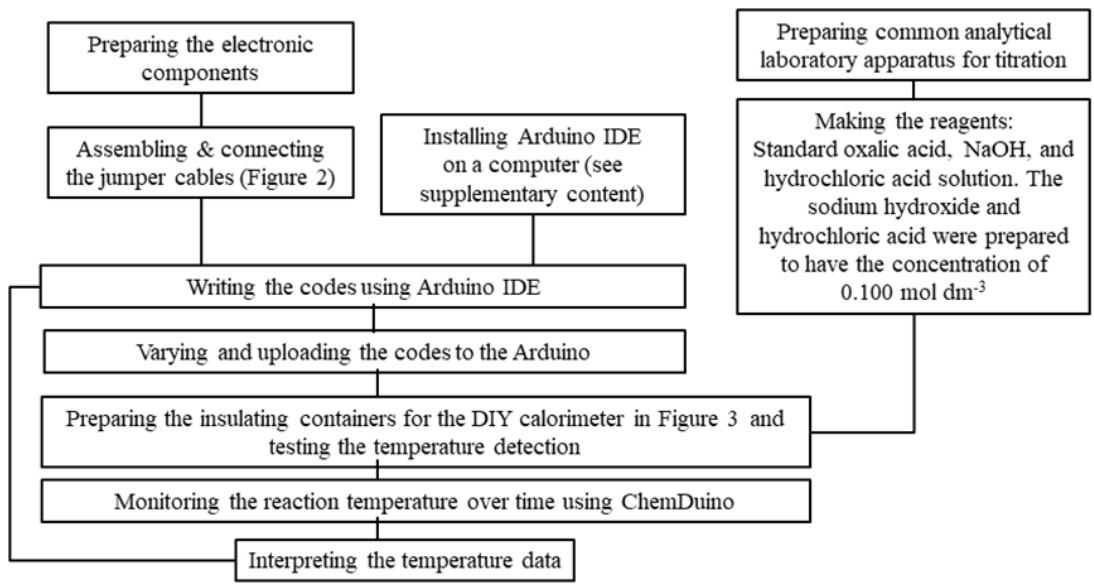

**FIGURE 1.** The overview of the steps used in this study

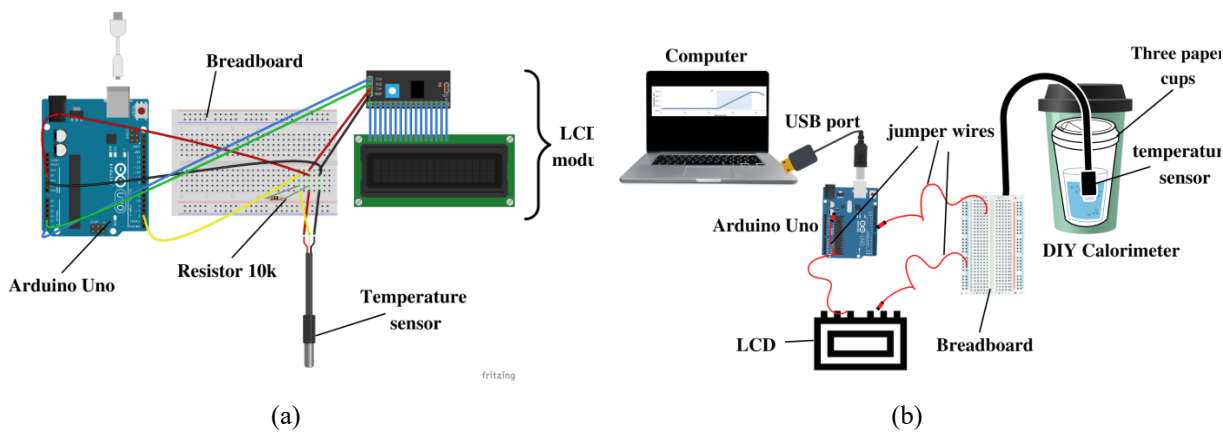

**FIGURE 2.** (a) The detailed schematic of the electrical circuit was made using Fritzing software version *0.9.3b* (b) The overview of the ChemDuino setup. The illustration was made using professional graphic design corel draw software and canva.

The method for fundamentally determining the enthalpy change of neutralization of sodium hydroxide and the hydrochloric acid solution is constant-pressure calorimetry. The device construction and experiment setup can be seen in Figure 2 (b). In advance, the ChemDuino was plugged into a laptop and simultaneously, the software Arduino IDE was turned on with the serial monitor opened. The initial temperatures of the two solutions were measured. Two 50-cm$^3$ burettes were used to deliver each 5.00 cm$^3$ of the acid and base solutions into the calorimeter, and the temperature changes were monitored from the LCD and computer screen. The experiment was repeated. The prototype was then evaluated, and the thermograms were analyzed. The enthalpy change of neutralization was calculated as follow [34]:

$$\Delta H_n = \frac{-m \times c \times \Delta T}{mol\ of\ water\ produced\ from\ the\ reaction}$$

Where m is the mass of water (in grams), c is the specific heat of water (4.182 JK$^{-1}$g$^{-1}$) and $\Delta T$ is the increase in temperature in Kelvin. $\Delta H_n$ unit is kJmol$^{-1}$.

# RESULTS AND DISCUSSION

## ChemDuino-Calorimeter Test Results with the Temperature Measurements

The prototype ChemDuino was tested to measure temperature change in the solution repeatedly. As a comparison, a standard analogue alcohol thermometer was used. The analogue thermometer can only measure to the nearest ±0.5°C, e.g. a temperature increase by 0.62°C cannot be detected. It is not possible to have in-situ temperature data acquisition. On the other hand, the DS18B20 sensor in ChemDuino recorded an increase of 0.62°C automatically to the computer. With the involvement of Arduino IDE software, the temperature in the solution was monitored in real-time on the computer screen via the serial monitor tool. With the implementation of green chemistry principles in the experimental procedures in which small amounts of volume and low concentration are used, the performance of the analogue thermometer can no longer keep up with the performance of the electronic sensor in ChemDuino. In Figure 3, the thermogram was conveniently produced using Excel. Alternatively, Excel spreadsheets can be used in tandem with PLX-DAQ (Parallax Data Acquisition) [35, 36]. The temperature display in ChemDuino can be programmed to show temperature directly into Kelvin and convert any temperature unit. The programmable ChemDuino is an advantage in calorimetry.

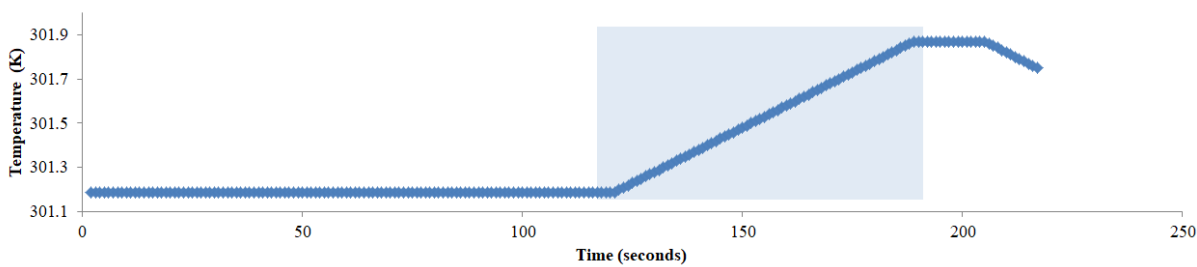

**FIGURE 3.** The thermogram shows time (in seconds) and temperature (in Kelvin); The data were brought together using ChemDuino code version 1.1, which is shown in supplementary content (3); data acquisition took 500 milliseconds as delay.

## ChemDuino-Calorimeter Design Results

The ChemDuino prototype is a pocket-sized device and lightweight. The total mass of the electronics is only around 50 grams. The temperature sensor is flexible and small enough to be inserted into the lid in Figure 2 (b). However, a waterproof-container design is needed to protect the electronic parts so that it is not too exposed. The smaller diameter paper cups are learnt to be beneficial. This is to ensure that the sensor's tip is submerged in the solution. Figure 4 describes the different codes used in developing the ChemDuino prototype. As shown in ChemDuino code version 1.0, the delay function is added to mimic the measurement of temperature changes using an analog thermometer (e.g., mercury or alcohol thermometer) under a specific time interval [37]. However, adding a more extended delay function in the codes is catastrophic within the experiments conducted with code version 1.0. This is because ChemDuino might lose the chance of detecting the maximum temperature in the exothermic neutralization reaction. In code version 1.0, the delay function is written delay (1000) in Figure 4 (a), which means the Arduino will remain idle for the 1000 milliseconds duration. Due to the smaller total volume and relatively low concentration used in the experiments, the maximum temperature change might be brief. In this study, most of the experiments were conducted using 5.00 $cm^3$ for each sodium hydroxide and hydrochloric acid solution to reduce the extensive use of chemicals in the laboratory. The solution concentration is also lowered from 1.00 mol $dm^{-3}$ (in a standard thermometric procedure) to 0.100 mol $dm^{-3}$ of both acid and base solution. With this condition, a delay function for 1000 milliseconds has a disadvantage in the code version 1.0. For this reason, the delay function was reduced to half in code version 1.1, as shown in Figure 4 (b). As a result, ChemDuino code version 1.1 is more sensitive to temperature fluctuation due to the *sensor.Set Resolution (12)* function, allowing Arduino to collect data to the nearest 0.0625°C. It also has an additional function directed to the measurement, e.g. conversion to other temperature units.

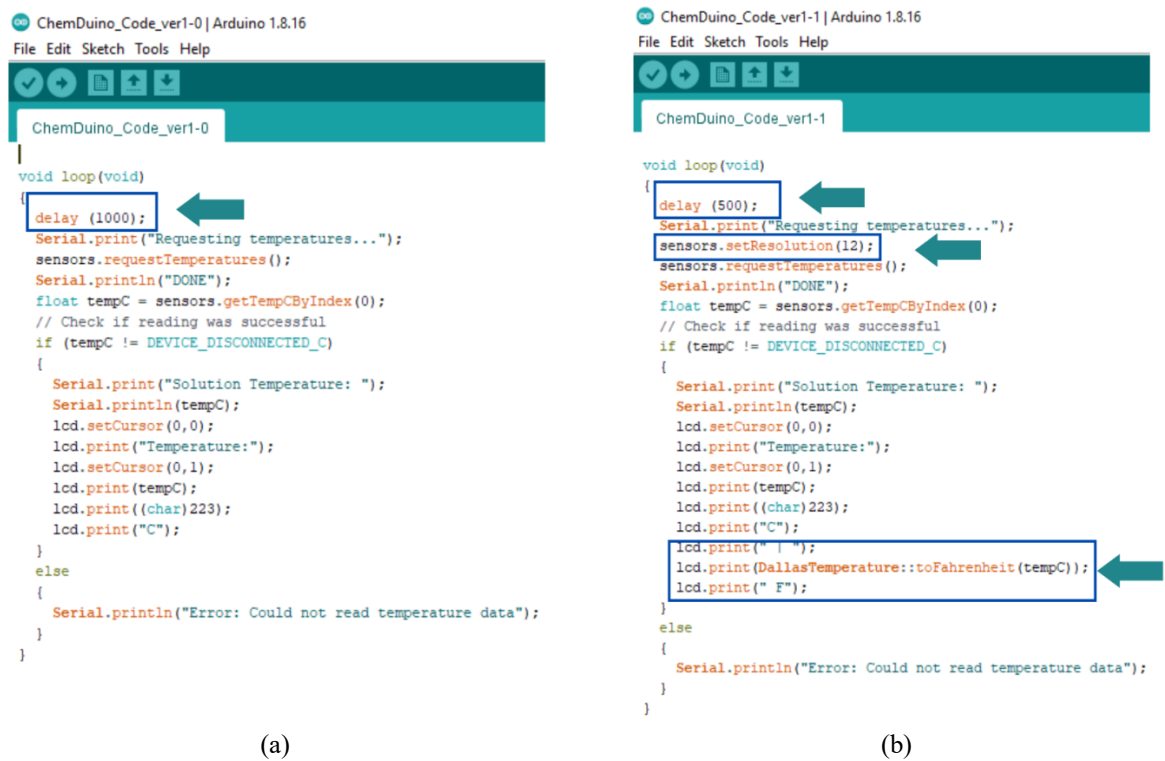

(a)                                                                 (b)

**FIGURE 4.** Comparing the two versions of sketch codes to data temperature acquisition, shown in (a) version 1.0 and (b) version 1.1 (see supplementary information).

### Analysis of the Enthalpy Change of Neutralization of a Strong Acid and Base Reaction

Table 1 shows the comparison results for the same measurement condition between ChemDuino with code version 1.0 and 1.1. The numbers of experiments repeated were the same. However, measurement with code version 1.0 gave a relatively higher standard deviation (± 5.2) than that found with code version 1.1 (± 1.9). The temperature data obtained from measurement with the older version is more spread out. The precision is improved with code version 1.1. The measurements with the code version 1.0 gave the average enthalpy change of neutralization to be −55.16 kJ mol$^{-1}$. However, with the code version 1.1, the average enthalpy change of neutralization was found to be −56.87 kJ mol$^{-1}$. The percentage differences with reference values are consistently reduced with the modification from code version 1.0 to version 1.1. In this study, a 50-cm$^3$ burette was used to measure 5.00 cm$^3$ of the solutions.

**TABLE 1.** Comparison Results of the Enthalpy of Neutralization of an Acid−Base Reaction Using Two Different Codes on ChemDuino and the Differences with the Reference Values

| Parameters | Average Values, kJ mol$^{-1}$ (± estimated standard deviation) | |
|---|---|---|
|  | Code version 1.0, N = 7[c] | Code version 1.1, N = 7[c] |
| ChemDuino Calorimeter | −55.16 (± 5.2) | −56.87 (± 1.9) |
| Difference, [a]% | 6.19 | 3.28 |
| Error [b]% | 3.45 | 0.455 |

[a]The percent difference shows the results between the ChemDuino and another similar experiment with paper cup calorimeter published by Bopegedera and Perera in 2017 (Instructor value −58.8 kJ mol$^{-1}$) [25]. [b]The percent error compares the experimental determination of the enthalpy change of neutralization of hydrochloric acid and sodium hydroxide solution with the theoretical value (−57.13 kJ mol$^{-1}$ at 298 K) [38]. [c]$N$ is the number of experiments.

In the future, the method can be further improved by minimizing the contributing factors in percentage error and uncertainties, e.g. using micro burette to measure the volume of the solutions.

The results of the temperature change data from the ChemDuino-calorimeter and conventional calorimeter were compared in Table 2. The experiments were conducted with 300 cm$^3$ sodium hydroxide 1.00 mol dm$^{-3}$ and 300 cm$^3$ hydrochloric acid 1.00 mol dm$^{-3}$ solutions. The volume and concentration values were similar to the previous research in 2020 [27]. The temperature increase was recorded in two significant figures. The experiments with the conventional method were done with a styrofoam cup, a plastic lid, and an alcohol thermometer (the smallest scale, 1.0°C). The experiments were repeated seven times, and the temperature data were recorded. The ChemDuino-calorimeter developed in this study was used as a comparison. Two-tailed independent Mann−Whitney U-tests were conducted to determine if the differences in these temperature change data from the two calorimeters. There was a statistical difference between the temperature change data measured with the ChemDuino and conventional calorimeter ($p<0.05$). The results indicated that the mean temperature change data using the developed ChemDuino calorimeter were higher than those recorded using the conventional calorimeter at a statistically significant level.

**TABLE 2.** Mann–Whitney U test results of the Temperature Change Data Measured with ChemDuino and Conventional Calorimeter

| Methods | N | Mean rank | Sum of ranks | Mann Whitney U | Z | p |
|---|---|---|---|---|---|---|
| ChemDuino[a] | 7 | 9.33 | 56 | 1 | 2.642 | 0.0083[b] |
| Conventional | 7 | 3.67 | 22 | | | |

[a]ChemDuino with the code version 1.1 was used. [b]The p-value is 0.0083 and the critical value of U at $p < 0.05$ is 5. The result is significant at $p < 0.05$.

# CONCLUSION

The prototype utilizing Arduino has been successfully developed and tested. It is evident that ChemDuino-Calorimeter can measure the temperature change from the neutralization reaction, even with a small-scale setup with relatively low volumes and concentrations. The total volume of acid and base solution used is just 10.00 cm$^3$ with a concentration of 0.100 mol dm$^{-3}$ for both of the solutions. In this study, working with an electronic sensor as an alternative tool allows us to reduce the chemical usage for the enthalpy change of neutralization measurements. This gives a "greener" experiment for the alternative implementation in thermochemistry. With the code version 1.0, the average enthalpy change of neutralization was found to be −55.16 kJ mol$^{−1}$, whereas, with version 1.1, the average enthalpy change of neutralization was found to be −56.87 kJ mol$^{−1}$. This yields an improvement in accuracy. The developed code version 1.1 provides faster data acquisition from the temperature sensor DS18B20. The precision of temperature detection is increased by adding a *sensor.Set Resolution (12)* function. This provides better data reproducibility, shown in a lower standard deviation. According to the Mann-Whitney U-test results, there is a statistically meaningful difference between the temperature change data measured with ChemDuino-Calorimeter and conventional calorimeter [U=1; $p < 0.05$]. In the future, this can be developed further in various thermochemistry-related experiments.

# ACKNOWLEDGMENTS

We would like to express our sincere gratitude to all the inspiring educators at the State University of Jakarta, Indonesia.